\documentclass[english,aps, twocolumn]{revtex4}
\usepackage[T1]{fontenc}
\usepackage[latin9]{inputenc}
\usepackage{color}
\usepackage{textcomp}
\usepackage{amssymb}
\usepackage{graphicx}
\usepackage{esint}

\makeatletter

\providecommand{\tabularnewline}{\\}

\@ifundefined{textcolor}{}
{%
 \definecolor{BLACK}{gray}{0}
 \definecolor{WHITE}{gray}{1}
 \definecolor{RED}{rgb}{1,0,0}
 \definecolor{GREEN}{rgb}{0,1,0}
 \definecolor{BLUE}{rgb}{0,0,1}
 \definecolor{CYAN}{cmyk}{1,0,0,0}
 \definecolor{MAGENTA}{cmyk}{0,1,0,0}
 \definecolor{YELLOW}{cmyk}{0,0,1,0}
 }

\makeatother

\usepackage{babel}
\begin{document}

\title{High energy Free Electron Laser for nuclear applications}

\author{R. Brinkmann$^{1}$, F. Camera$^{2,3}$, I. Drebot$^{3}$, V. Petrillo$^{2,3}$,
M. Rossetti Conti$^{2,3}$, L. Serafini$^{3}$}

\affiliation{$^{1}$ Deutsches Elektronen-Synchrotron DESY, Notkestraße 85, 22607
Hamburg, Germany}

\affiliation{$^{2}$Università degli Studi di Milano, via Celoria 16, 20133, Milano,
Italy}

\affiliation{$^{3}$INFN-Sezione di Milano, via Celoria 16, 20133, Milano, Italy}
\begin{abstract}
The capability of Free-Electron Lasers to generate photon beams with
record performances in the domain of MeV-class photon energy for nuclear
photonics applications is here analyzed. We discuss possible nuclear
FEL working points. Some typical situations are presented and simulated
by means of the three-dimensional code GENESIS 1.3. The losses by
spontaneous emission and by wakefield induced energy spread are then
evaluated. The calculations show the possibility of generating FEL
radiation at 1-4 MeV with less than 300 eV of bandwidth with not prohibitive
set up requirements, achieving therefore regimes where no other kind
of gamma ray sources could arrive.
\end{abstract}
\maketitle

\section{Introduction}

Present X-ray Free-Electron Lasers (FELs) in operation or under commissioning
achieve maximum photon energies between 10-20 keV \cite{LCLS,SCSS,XFEL,PSI},
a range covering an extremely wide number of applications in the imaging
field and in the atomic and electronic physics. For larger photon
energies, entering therefore the Nuclear Photonics domain, which starts
at around 1 MeV and extends up to several MeVs, the radiation sources
of reference at the state of art and for the near future developments
are, at this moment, the Inverse Compton Scattering sources (ICSs)\cite{HIGS,ELI}.
ICSs are based on the spontaneous incoherent radiation generated by
Compton back-scattering of lasers by high brightness electron beams
\cite{HIGS,ELI} and typically are projected to provide few $10^{8}$
polarized photons at 1-20 MeV in a bandwidth not narrower than several
$10^{-3}$, namely 5-100 keV absolute bandwidth, with a completely
incoherent radiation structure. 

It has been debated whether a scientific case exists for FELs in this
regime. Despite the large costs inherent in the development of Linac
based FELs with electron beam in the range $20-50$ GeV, a strong
scientific case justifies a design study within the frame of a Linear
Collider scenario, i.e. conceiving a parasitic use for a nuclear oriented
FEL of a separate beam line at a TeV-class Linac based collider. Using
a superconducting linac \cite{XFEL} in re-circulating mode by means
of a return loop to double the energy to >30 GeV could also be an
option. Although FELs were born and developed mainly with the goal
to advance scientific researches and knowledge in the science of matter
and related fields, by delivering the most advanced coherent and bright
X-ray photon beams, it is worthwhile to consider exploiting their
extreme capabilities to generate photon beams with record performances
also in the domain of MeV-class photon energy. The coherent and cooperative
mechanism for emitting radiation proper of FELs allows for a large
upgrade of the photon beam properties if the FEL collective instability
can be operated up to such a high photon energy. Together with coherence
and enhancement in the radiated power goes an extremely narrow bandwidth
proper of FELs operation, approaching the range of O($10^{-4}$) for
the rms relative bandwidth, compared to the aforementioned O($10^{-3}$)
of ICSs. This is extremely crucial for Nuclear Photonics applications,
in which bandwidth, as well as spectral density, are of the maximum
relevance. The achievement of absolute bandwidths down to the 100
eV range (relative bandwidth smaller than O($10^{-4}$) into the O($10^{-5}$)
range) would open a completely new approach in Nuclear Photonics,
enabling the excitation of single nuclear states, now forbidden to
ICSs since their best performance is limited to more than two order
of magnitude above this value. 100 eV absolute bandwidth would therefore
represent a true breakthrough in Nuclear Physics and Photonics \textendash{}
the best energy resolution achieved with pure Ge detectors and close
to the intrinsic single nuclear state bandwidth \cite{nucl}. The
next generation ICSs will probably be able to go down by one order
of magnitude in relative bandwidth, approaching 1 keV, still far from
the 100 eV level, the main issue being the collision laser intrinsic
bandwidth, unlikely to become narrower than $5\,10^{-4}$. The real
challenge for nuclear oriented FELs will be the capability to operate
below the O($10^{-4}$) bandwidth, implying severe requests to the
accuracy and precision of undulator field stability and uniformity,
as well as extremely low emittances and energy spreads of the electron
beam.

In this paper, first we discuss the possible nuclear FEL working points.
Then, some characteristic situations are presented and simulated by
means of the FEL code{\small{} GENESIS 1.3} \cite{genesis}. The losses
by spontaneous emission and by wakefield induced energy spread are
then evaluated. We close with comments and conclusions.

\section{Working point discussion}

The aim is to produce FEL radiation with energy $E_{phot}\,\,(E_{phot}(\mathrm{MeV})=1.25\,10^{-6}/\lambda(\mu\mathrm{m}))$
larger than $1\,\mathrm{MeV}$, with absolute bandwidth approaching
$100\mathrm{\, eV}$ ($\delta E(\mathrm{MeV})=\beta10^{-4}$), with
the least demanding requirements for undulator, electron beam current
and energy.

The Pierce parameter \cite{key-1}, describing the one-dimensional
FEL dynamics, can be expressed as:

\begin{equation}
\rho=5.7\,10^{-3}\frac{J^{1/3}}{\gamma}(a_{w}\lambda_{w})^{2/3}=\frac{\beta10^{-4}}{E_{phot}}\label{eq:1}
\end{equation}
where $\lambda_{w},$ here in micron, is the undulator period, $a_{w}$
the undulator parameter and $J=I/\sigma_{0}^{2}(\mathrm{A/\mu m^{2})}$
and gives an estimate of the natural FEL bandwidth $\Delta\lambda/\lambda\approx\rho.$
Furthermore,

\begin{equation}
\lambda=\frac{1.25\,10^{-6}}{E_{phot}}=\frac{\lambda_{w}}{2\gamma^{2}}(1+a_{w}^{2})\label{eq:2}
\end{equation}
is the radiation resonant wavelength.

By combining (\ref{eq:1}) and (\ref{eq:2}) and considering the linear
relation between undulator period and strength $a_{w}=6.57\,10^{-5}B(\mathrm{T})\lambda_{w}(\mu\mathrm{m})=\alpha\lambda_{w}(\mu\mathrm{m})$
(being therefore $\alpha=6.57\,10^{-5}B(\mathrm{T})\,\mu\mathrm{m}^{-1}$
with $B$ peak field of the magnets in a planar undulator), the nuclear
FEL conditions can be written as:

\begin{equation}
\gamma=7\,10^{3}\frac{\beta(1+(\alpha\lambda_{w})^{2})}{J^{1/3}\alpha^{2/3}\lambda_{w}^{1/3}},
\end{equation}
with a corresponding photon energy:

\begin{equation}
E_{phot}=\frac{1.22\,10^{2}\beta^{2}(1+(\alpha\lambda_{w})^{2})}{J^{2/3}\alpha^{4/3}\lambda_{w}^{5/3}}.
\end{equation}
For $\beta=3$ ($\delta E=300\,\mathrm{eV}$) and $\beta=5$ ($\delta E=500\,\mathrm{eV}$),
graphs of $\gamma$, $E_{phot}$ and $\rho$ as functions of $\lambda_{w}$
(whose state of the art is so far around half centimeter with $a_{w}\ll1$
\cite{und corti,und corti2}) for $J=2000$ and $1000$ and $\alpha=1.4\,10^{-4}\,\mu\mathrm{m}^{-1}$
are presented in Fig. \ref{fig:-(solid-line),}. 

\begin{figure}
\includegraphics[width=1\columnwidth]{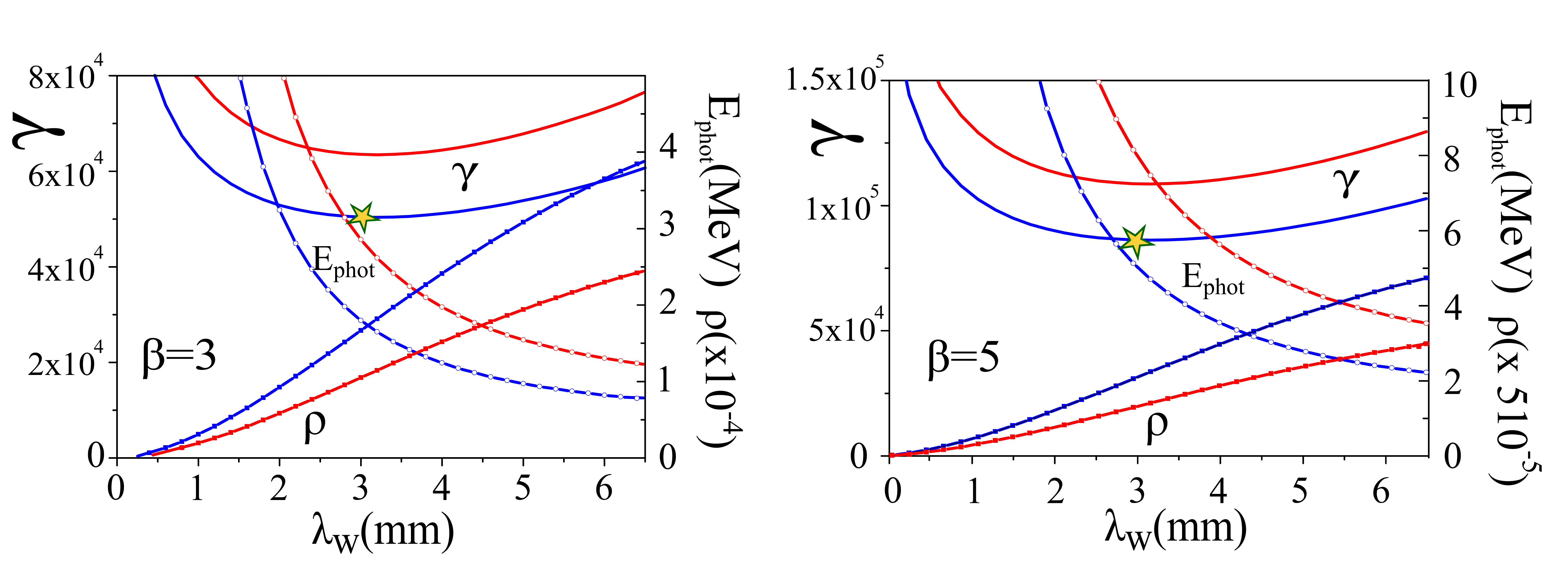}

\caption{\label{fig:-(solid-line),}$\gamma$ (solid line), $E_{phot}$ (solid
line with open circles) and $\rho$ (solid line with squared) as function
of $\lambda_{w}$ for $J=2000\, A/\mu m^{2}$ (blue), $J=1000\, A/\mu m^{2}$
(red), for (left)$\beta=3$($\delta E=300\,\mathrm{eV}$), (right)
$\beta=5$ ($\delta E=500\,\mathrm{eV}$),$\alpha=1.47\,10^{-4}$.}
\end{figure}

As can be seen, while the photon energy decreases monotonically, $\gamma$
passes through a flat minimum at $\lambda_{w}=1/(\sqrt{5}\alpha)=3.2\,\mathrm{mm}$
whose value is:

\begin{equation}
\gamma=1.1\,10^{4}\frac{\beta}{(\alpha J)^{1/3}},\label{eq:5}
\end{equation}

with a corresponding photon energy of:
\begin{equation}
E_{phot}=\frac{5.59\,10^{2}\beta^{2}\alpha^{1/3}}{J^{2/3}}.\label{eq:6}
\end{equation}

Photon energies (> 1 MeV) and bandwidths ($\beta$ approaching 1)
suitable for nuclear applications are therefore obtained with current
densities $J$ in the range:

\begin{equation}
J<1.32\,10^{4}\beta^{3}\alpha^{1/2},
\end{equation}

and corresponding Pierce parameters:

\begin{equation}
\rho<2.9\,10^{-4}\beta(\alpha\lambda_{w})^{4/3}.
\end{equation}

\begin{figure}
\includegraphics[width=1\columnwidth]{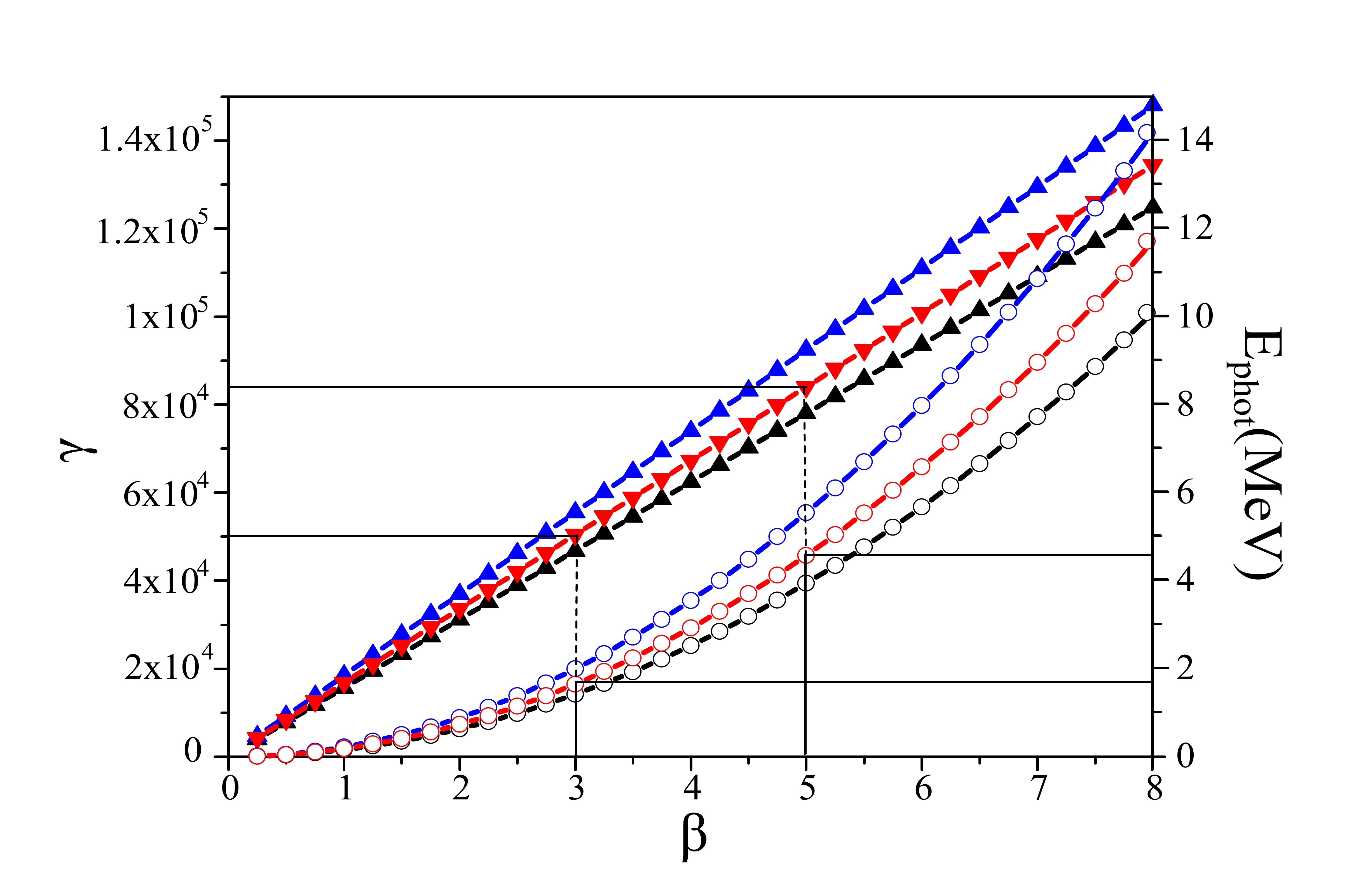}\caption{\label{fig:gamma-and-Ephot}$\gamma$ and $E_{phot}$ as function
of $\beta$ for: red $J=2000\, A/\mu m^{2}$, black $J=2500\, A/\mu m^{2}$,
blue $J=1500\, A/\mu m^{2}$}

\end{figure}

Eq.s \ref{eq:5} and \ref{eq:6} are shown in Fig \ref{fig:gamma-and-Ephot}.

Values of $\lambda_{w}$ around or slightly below the minimum of $\gamma$
identify useful working points. Relying on Fig.\ref{fig:-(solid-line),},
lower periods lead to a favorable trend of the photon energy, but
worse conditions in $\rho$ and $\gamma$, corresponding to longer
undulators and less radiation photons. Going towards longer periods,
instead, the $\rho$ parameter increases, but the energy of the radiation
photons tends to go below the MeV level. 

For $\beta=3$ ($\delta E=300\,\mathrm{eV}$), situations with $E_{phot}>1$
MeV can be obtained with $\lambda_{w}\approx3$ mm and $\gamma\gtrsim5$
$10^{4}$. With J=2000 $\mathrm{A}/\mu\mathrm{m}^{2}$, the $\rho$
parameter turns out to be about $1.5\,10^{-4}$. It is possible to
achieve similar bandwidth levels using lower current densities J=1000
$\mathrm{A}/\mu\mathrm{m}^{2}$, together with larger electron energies
$\gamma\gtrsim6.5$ $10^{4}$, with consequent increase in the photon
energy $E_{phot}\approx4$ MeV. In this case the FEL parameter does
not exceed $10^{-4}$, leading to longer gain and saturation lengths
and smaller gains. The case $\beta=5$ is characterized by larger
electron and radiation photon energy with a consequent smaller $\rho$,
longer undulators and lower radiation yield. 

These two working points are presented in Table 1. 
\begin{table}
\begin{tabular}{|c|c|c|}
\hline 
 & case A & case B\tabularnewline
\hline 
\hline 
$E_{_{phot}}(\mathrm{MeV})$ & 1.8 & 4.3\tabularnewline
\hline 
$\Delta E_{phot}(eV)$ & 300 & 500\tabularnewline
\hline 
$\Delta E_{phot}/E_{phot}$ & $1.6\,10^{-4}$ & $1.1\,10^{-4}$\tabularnewline
\hline 
E (GeV) & 25  & 40\tabularnewline
\hline 
I (kA) & 12.5 & 12.5\tabularnewline
\hline 
$\sigma_{x,y}(\mu\mathbf{\mathrm{m}})$ & 2.5 & 2.5\tabularnewline
\hline 
$\varepsilon(\mathrm{mm\, mrad})$ & 5~10$^{-3}$-4~10$^{-2}$ & 5~10$^{-3}$-2.5~10$^{-2}$\tabularnewline
\hline 
$\triangle E/E$ & 0.5~10$^{-4}$-10$^{-4}$ & 0.5~10$^{-4}$-10$^{-4}$\tabularnewline
\hline 
$\lambda_{w}(\mathrm{mm})$ & 3 & 3\tabularnewline
\hline 
$a_{w}$ & 0.42 & 0.42\tabularnewline
\hline 
$\rho$ & 1.710$^{-4}$ & 1.3 10$^{-4}$\tabularnewline
\hline 
\end{tabular}

\caption{Electron, undulator and photon parameters of two possible working
points.}
\end{table}
Regarding their feasibility, the state of the art of the undulators
is so far at 4 mm of period with $B\approx0.5$ T, using the cryogenic
technology \cite{und corti,und corti2}. We are exploring therefore
a regime not much far from this one. A superconducting helical undulator
could be used to further reduce the gain length. The requirements
on the electron beam will be commented in the next Section.

\section{Three-dimensional simulations}

Three-dimensional time-dependent simulations have been performed with
the code {\small GENESIS 1.3} \cite{genesis} for both working points
described in Table 1, with varying electron emittance and energy spread
and using the best numerical accuracy options. In these simulations,
due to the low emittance and to the relatively short saturation length,
the electron beam transverse dimensions have not be taken under control
by any external focusing system. They increase therefore along the
undulator from 2.5 $\mu m$ to about 5 $\mu m$ in the worst case,
leading to slightly longer gain lengths and lower saturation power
levels with respect to those evaluated with the initial values, but
also slightly smaller bandwidths in those cases that are not strongly
dominated by emittance and energy spread.

The results of the simulations are summarized in Fig. 3 (case A) and
Fig. 4 (case B), where the bandwidth (a) and the number of photons
(b) are reported as function of the emittance for two values of energy
spread.

\begin{figure}
\includegraphics[width=1\columnwidth]{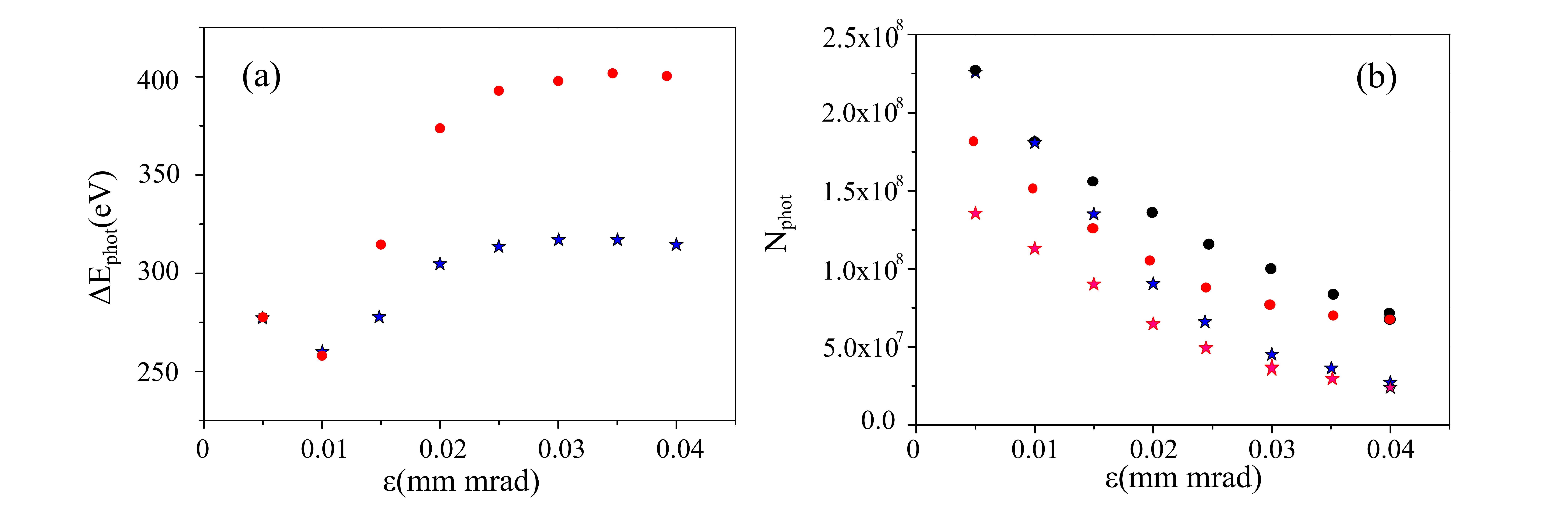}

\caption{Case A. (a) Blue stars $\triangle E/E=0.5\,10^{-4}$, red circles
$\triangle E/E=\,10^{-4}$. (b) Blue circles $\triangle E/E=0.5\,10^{-4}$,
L=20 m, blue stars $\triangle E/E=0.5\,10^{-4}$, L=30 m, red circles
$\triangle E/E=\,10^{-4}$ L=20 m, red stars $\triangle E/E=\,10^{-4}$
L=30 m.}

\end{figure}

In case A, ($\rho=1.4\,10^{-4},\triangle E/E=0.5\,10^{-4},\,10^{-4}<\rho)$,
the bandwidth at very low emittance assumes values close to its one-dimensional
natural value $\rho E_{phot}$, and then increases with the emittance
according to the trend given by the Ming Xie formulas \cite{key-2}.
At the same time, the number of photons decreases from few $10^{8}$
to less than $10^{7}$. The saturation length increases from 10-15
m at low emittance up to values larger than 30 m. In these cases we
have reported in the graph and in the Table the value of the energy
and number of photons at 30 m.

\begin{figure}
\includegraphics[width=1\columnwidth]{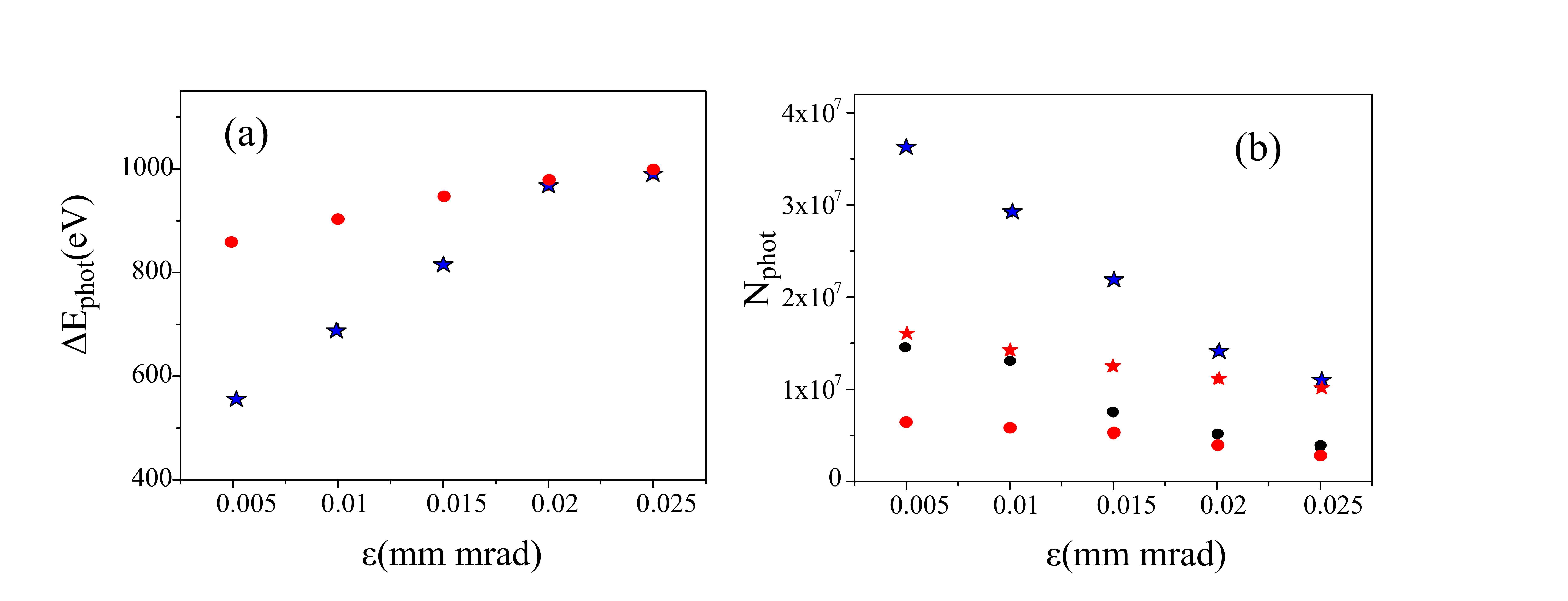}

\caption{Case B. (a) Blue stars $\triangle E/E=0.5\,10^{-4}$, red circles
$\triangle E/E=\,10^{-4}$. (b) Blue circles $\triangle E/E=0.5\,10^{-4}$,
L=20 m, blue stars $\triangle E/E=0.5\,10^{-4}$, L=30 m, red circles
$\triangle E/E=\,10^{-4}$ , L=20 m, red stars $\triangle E/E=\,10^{-4}$
, L=30 m.}

\end{figure}
With further larger emittance, since the electron beam transverse
dimension along the undulator increases with the emittance, the $\rho$
parameter becomes in average smaller, partially compensating the enlargement
due to three-dimensional effects. The bandwidth therefore presents
a plateau, with a subsequent decrease at increasing emittance. In
case B, ($\rho=0.95\,10^{-4},\triangle E/E=0.5\,10^{-4},\,10^{-4}\gtrsim\rho)$
at null emittance the bandwidth shows a significant influence of the
energy spread. 

\begin{table*}
\begin{tabular}{|c|c|c|c|}
\hline 
e-beam energy  & 0.31 GeV & absolute bandwidth  & 17.5 keV \tabularnewline
\hline 
Period & 0.5 \textmu{}m & norm. emitt. en (\textmu{}m)  & 0.5 \tabularnewline
\hline 
$a_{0}$ & 0.02  & energy spread $10^{-4}$ & 7. \tabularnewline
\hline 
Photon energy  & 3.5 MeV  & repetition rate  & 3.2 kHz \tabularnewline
\hline 
peak current  & 200 A  & average beam current  & 800 nA \tabularnewline
\hline 
e- bunch charge  & 250 pC  & average beam power  & 240 W \tabularnewline
\hline 
Laser energy & 0.2 J  & \# photons/pulse  & $1.1\,10^{5}$\tabularnewline
\hline 
e- beam spot size  & 20 \textmu{}m  & spectral density (s eV)$^{-1}$ & 7.6 10$^{3}$ \tabularnewline
\hline 
relative bandwidth  & 5.0 $10^{-3}$  & div. angle  & 40 \textmu{}rad \tabularnewline
\hline 
\end{tabular}\caption{ELI-NP-GBS parameters and expected performances.}
\end{table*}

The absolute bandwidth is more than one order of magnitude narrower
than ELI-NP-GBS (see Table 2), the number of photon per pulse is three
order of magnitude larger, leading to spectral densities four order
of magnitude higher. In Table 3, the simulation parameters and results
of the FEL source are listed. As can be seen, at 20 pC, levels of
electron emittance up to 4 $10^{-2}$ mm mrad can be tolerated with
energy spreads 0.5-1 $10^{-4}$, not too far from the state of the
art reaching $10^{-1}$ mm mrad and $10^{-4}$ relative energy spread.
Better beam quality requires further improvements in high-brilliance
beam sources, which may also include plasma-wakefield driven injectors.
In order to decrease further the bandwidth we need to relax the undulator
parameter and/or the electron beam peak current (from tens kA down
to few kA): this increases substantially the technological challenges
on the beam quality (energy spread in particular) and undulator field
quality and uniformity. 

\begin{table*}
\textcolor{black}{}%
\begin{tabular}{|l|l|c|c|c|c|c|c|c|}
\hline 
 & \multicolumn{1}{l}{\textcolor{black}{case A}} & \multicolumn{1}{c}{} & \multicolumn{1}{c}{} &  & \multicolumn{1}{c}{\textcolor{black}{case B}} & \multicolumn{1}{c}{} & \multicolumn{1}{c}{} & \tabularnewline
\hline 
\hline 
\textcolor{black}{e- beam energy (GeV)} & \multicolumn{1}{l}{\textcolor{black}{25 }} & \multicolumn{1}{c}{} & \multicolumn{1}{c}{} &  & \multicolumn{1}{c}{\textcolor{black}{40 }} & \multicolumn{1}{c}{} & \multicolumn{1}{c}{} & \tabularnewline
\hline 
\textcolor{black}{Period (mm)} & \multicolumn{1}{l}{\textcolor{black}{~3 }} & \multicolumn{1}{c}{} & \multicolumn{1}{c}{} &  & \multicolumn{1}{c}{\textcolor{black}{3 }} & \multicolumn{1}{c}{} & \multicolumn{1}{c}{} & \tabularnewline
\hline 
\textcolor{black}{$a_{w}$} & \multicolumn{1}{l}{\textcolor{black}{~~0.35 }} & \multicolumn{1}{c}{} & \multicolumn{1}{c}{} &  & \multicolumn{1}{c}{\textcolor{black}{0.35 }} & \multicolumn{1}{c}{} & \multicolumn{1}{c}{} & \tabularnewline
\hline 
\textcolor{black}{e- beam charge (pC) } & \multicolumn{1}{l}{\textcolor{black}{20 }} & \multicolumn{1}{c}{} & \multicolumn{1}{c}{} &  & \multicolumn{1}{c}{\textcolor{black}{20 }} & \multicolumn{1}{c}{} & \multicolumn{1}{c}{} & \tabularnewline
\hline 
\textcolor{black}{e- beam length rms (fs) } & \multicolumn{1}{l}{\textcolor{black}{1 }} & \multicolumn{1}{c}{} & \multicolumn{1}{c}{} &  & \multicolumn{1}{c}{\textcolor{black}{1}} & \multicolumn{1}{c}{} & \multicolumn{1}{c}{} & \tabularnewline
\hline 
\textcolor{black}{e- beam initial spot size (\textmu{}m) } & \multicolumn{1}{l}{\textcolor{black}{~2.5 }} & \multicolumn{1}{c}{} & \multicolumn{1}{c}{} &  & \multicolumn{1}{c}{\textcolor{black}{2.5}} & \multicolumn{1}{c}{} & \multicolumn{1}{c}{} & \tabularnewline
\hline 
\textcolor{black}{Photon energy (MeV) } & \multicolumn{1}{l}{\textcolor{black}{1.85 ~~~}} & \multicolumn{1}{c}{} & \multicolumn{1}{c}{} &  & \multicolumn{1}{c}{\textcolor{black}{4.2~~}} & \multicolumn{1}{c}{} & \multicolumn{1}{c}{} & \tabularnewline
\hline 
\textcolor{black}{peak current (kA)} & \textcolor{black}{12.5  } & \textcolor{black}{12.5 } & \textcolor{black}{12.5 } & \textcolor{black}{12.5} & \textcolor{black}{12.5  } & \textcolor{black}{12.5 } & \textcolor{black}{12.5 } & \textcolor{black}{12.5}\tabularnewline
\hline 
\textcolor{black}{$\varepsilon_{n,slice}$(\textmu{}m) } & \textcolor{black}{0.01 } & \textcolor{black}{0.04} & \textcolor{black}{0.01} & \textcolor{black}{0.04} & \textcolor{black}{0.01 } & \textcolor{black}{0.025} & \textcolor{black}{0.01} & \textcolor{black}{0.025}\tabularnewline
\hline 
\textcolor{black}{slice energy spread x$10^{-4}$} & \textcolor{black}{0.5 } & \textcolor{black}{0.5} & \textcolor{black}{1} & \textcolor{black}{1} & 0.5 & 0.5  & 1  & 1\tabularnewline
\hline 
\textcolor{black}{$\rho$ initial x$10^{-4}$} & \textcolor{black}{1.5} & \textcolor{black}{1.5} & \textcolor{black}{1.5} & \textcolor{black}{1.5} & 0.95  & 0.95 & 0.95 & 0.95\tabularnewline
\hline 
\textcolor{black}{$\rho$ 3d initial x$10^{-4}$} & \textcolor{black}{1.2} & \textcolor{black}{0.47} & \textcolor{black}{0.95} & \textcolor{black}{0.4} & 0.7 & 0.45 & 0.42 & 0.3\tabularnewline
\hline 
\textcolor{black}{saturation length (m)} & \textcolor{black}{15} & \textcolor{black}{>30} & \textcolor{black}{22} & \textcolor{black}{>30} & \textcolor{black}{24 } & \textcolor{black}{>30} & \textcolor{black}{>30} & \textcolor{black}{>30}\tabularnewline
\hline 
\textcolor{black}{relative bandwidth x$10^{-4}$} & \textcolor{black}{1.4} & \textcolor{black}{1.7} & \textcolor{black}{1.4} & \textcolor{black}{2.1} & \textcolor{black}{1.65} & \textcolor{black}{2.3} & \textcolor{black}{2.1} & \textcolor{black}{2.3}\tabularnewline
\hline 
\textcolor{black}{absolute bandwidth (eV)} & \textcolor{black}{260} & \textcolor{black}{320} & \textcolor{black}{260} & \textcolor{black}{400} & \textcolor{black}{690} & \textcolor{black}{990} & \textcolor{black}{900} & \textcolor{black}{990}\tabularnewline
\hline 
\textcolor{black}{rad. energy/pulse (J) $\times10^{-6}$} & \textcolor{black}{21} & \textcolor{black}{$8^{*}$} & \textcolor{black}{17} & \textcolor{black}{$7.8^{*}$} & \textcolor{black}{26} & \textcolor{black}{3$^{*}$} & \textcolor{black}{17$^{*}$} & \textcolor{black}{2.8$^{*}$}\tabularnewline
\hline 
\textcolor{black}{\# photons/pulse $\times10^{7}$} & \textcolor{black}{18.3 } & \textcolor{black}{$7^{*}$} & \textcolor{black}{15.} & \textcolor{black}{$6.7^{*}$} & \textcolor{black}{10. } & \textcolor{black}{$1.2^{*}$} & \textcolor{black}{$6.8^{*}$} & \textcolor{black}{$1.1^{*}$}\tabularnewline
\hline 
\textcolor{black}{Spectral Density $(s\cdot eV){}^{-1}\times10^{7}$} & \textcolor{black}{2.8} & \textcolor{black}{0.9} & \textcolor{black}{2.3} & \textcolor{black}{0.67} & \textcolor{black}{1.6 } & \textcolor{black}{0.048} & \textcolor{black}{0.3} & \textcolor{black}{0.04}\tabularnewline
\hline 
\textcolor{black}{phot. beam div. angle ($\mu$rad)} & \textcolor{black}{1} & \textcolor{black}{0.78 } & \textcolor{black}{0.8} & \textcolor{black}{0.7} & \textcolor{black}{0.28} & \textcolor{black}{0.25} & \textcolor{black}{0.42} & \textcolor{black}{0.35}\tabularnewline
\hline 
\textcolor{black}{radiation size($\mu$m)} & \textcolor{black}{5 } & \textcolor{black}{3} & \textcolor{black}{5.5} & \textcolor{black}{3.5} & \textcolor{black}{4} & \textcolor{black}{3.5} & \textcolor{black}{5} & \textcolor{black}{4}\tabularnewline
\hline 
\textcolor{black}{Current limit for losses (A)} & \textcolor{black}{1650} &  &  &  & \textcolor{black}{2240 } &  &  & \tabularnewline
\hline 
\textcolor{black}{incoh. div. angle (\textmu{}rad) } & \textcolor{black}{25 } &  &  &  & \textcolor{black}{10 } &  &  & \tabularnewline
\hline 
\end{tabular}\caption{FEL simulation with GENESIS 1.3. The symbol {*} indicates that the
estimates have been done at 30 m. }
\end{table*}

In Fig 5 we show the power growth as function of the undulator length,
for case A with emittance $\varepsilon=0.01\,\mathrm{\mu m}$ and
two values of energy spread. Saturation is achieved in 17-20 m, with
power levels of 1-10 GW. The bandwidth achieves 260 eV, a bit less
than the nominal value of 300 eV. The photon beam divergence angle
is dominated by the electron dynamics. 

\begin{figure}

\includegraphics[width=0.9\columnwidth]{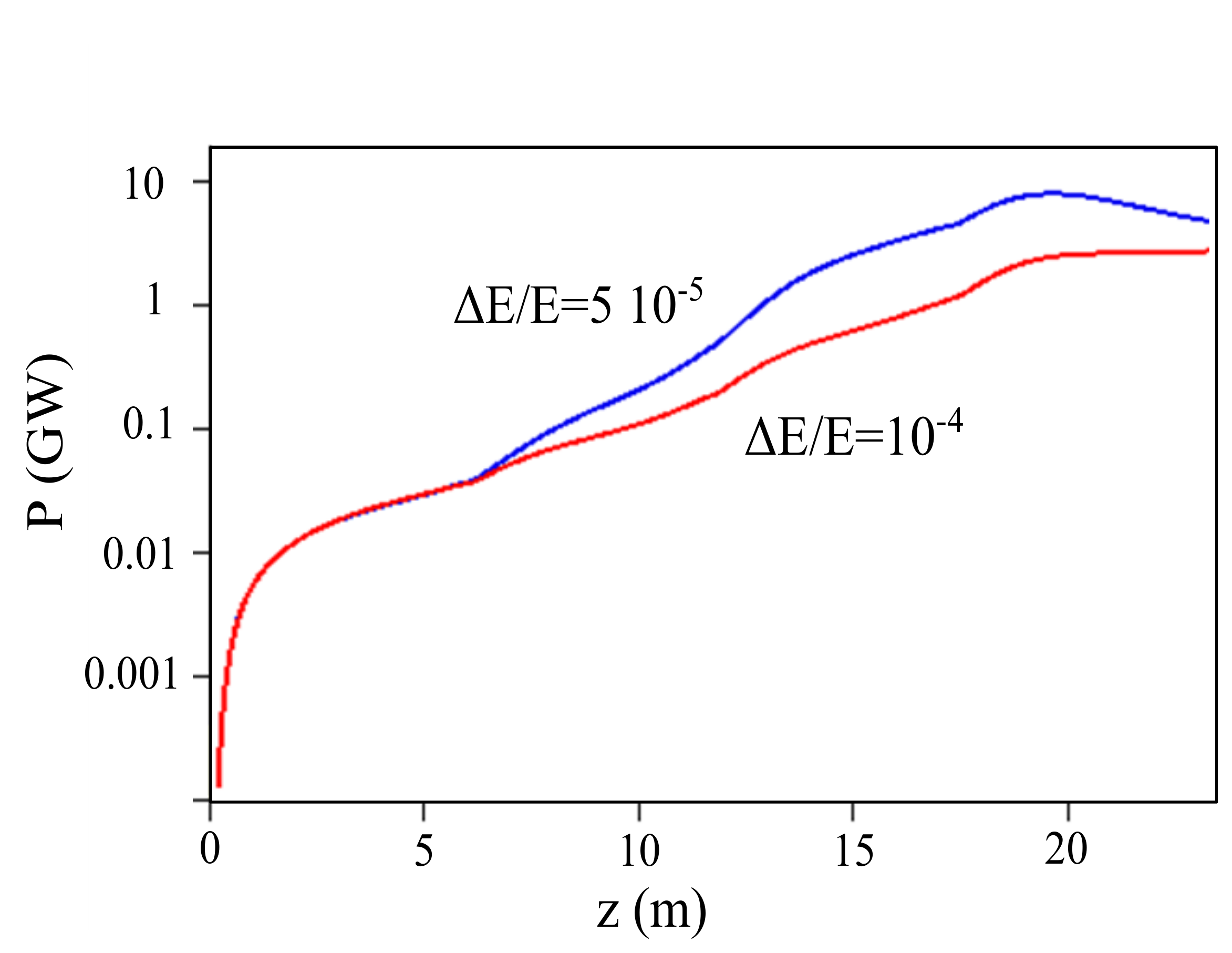}\caption{Power in GW as function of the undulator length in m for case A and
two values of energy spread.}

\end{figure}

\section{Spontaneous undulator emission and wakefield induced energy spread
losses}

Possible sources of losses are the spontaneous undulator emission
in the first undulator stages and the wakefiels induced energy spread.
If the energy losses due to undulator spontaneous emission in the
first FEL gain lengths are intense, the FEL process does not start.
The spectral broadening due to the spontaneous emission can be estimated
through a corresponding induced electron energy spread given by: 
\[
\left[\Delta\gamma^{2}\right]_{loss}=\frac{14}{15}(2\pi)^{2}\frac{r_{e}\lambda_{c}}{\lambda_{w}^{3}}a_{w}^{2}L_{w}\gamma^{4}
\]
 as derived in the Appendix, in agreement with the result of Ref {[}LCLS{]}.

The coherent emission can occur only if the relative bandwidth enlargement
due to this energy spread within the first gain length $L_{G}=\lambda_{w}/(4\pi\sqrt{3}\rho)$
is less than the natural FEL bandwidth $\rho$: 
\[
\left[\frac{\Delta\gamma^{2}}{\gamma^{2}}\right]_{loss}=\frac{14}{15}(2\pi)^{2}\frac{r_{e}\lambda_{c}}{\lambda_{w}^{3}}a_{w}^{2}L_{G}\gamma^{2}<\rho^{2}
\]

This can be translated into a condition for the current density:

\[
J(\mathrm{A/\mu m^{2}})>3.5\,10^{-20}\gamma_{e}^{5}/\lambda_{w}^{4}.
\]

In our cases, for both working points, the density current limits
are lower than the value chosen. 

As regards the electron energy spread induced by the wakefields, we
have evaluated numerically the distortion in the electron distribution
due to the wakefields during the propagation along the undulator,
assuming a copper beam pipe with millimeter size and electron beams
with the characteristics of the working points analyzed. Since the
total charge is very small and the electron beam very short, the wakefield
effect is completely negligible.

\section{Conclusions}

The capability of Free-Electron Lasers to generate photon beams with
record performances in the domain of MeV-class photon energy for nuclear
photonics applications is here analyzed. We discuss possible nuclear
FEL working points. Some typical situations are presented and simulated
by means of {\small GENESIS 1.3}. The losses by spontaneous emission
and by wakefield induced energy spread are then evaluated.

\section*{Appendix}

Recalling the definition of undulator parameter $a_{w}=\frac{eB_{0}}{\sqrt{2}mc^{2}k_{w}}$,
the total energy of the undulator field $E_{w}=\int dVw=\frac{\Sigma L_{w}}{8\pi}B_{0}^{2}$
($w=\frac{1}{8\pi}B_{0}^{2}$ being the magnetic energy density, $\Sigma$
the transverse area, $k_{w}$ the undulator period and $L_{w}$ the
undulator length), is:
\begin{equation}
E_{w}=\frac{a_{w}^{2}\Sigma L_{w}m^{2}c^{4}k_{w}^{2}}{4\pi e^{2}}.
\end{equation}
 The total photon number emitted in the spontaneous undulator radiation
process can be estimated through the luminosity formula as:

\begin{equation}
N_{phot}=\sigma_{T}\frac{N_{e}N_{w}}{\Sigma}
\end{equation}
where $\sigma_{T}=\frac{8\pi}{3}r_{e}^{2}$ with $r_{e}=e^{2}/(mc^{2})=2.81$
$10^{-15}$ m the classical electron radius, $N_{e}$ number of electron
and $N_{w}=E_{w}/\hbar ck_{w}$ number of the pseudophotons of the
undulator.

Collecting (8) and (9), we have:

\[
N_{phot}=\frac{2}{3}e^{2}\frac{N_{e}a_{w}^{2}L_{w}k_{w}^{2}}{\hbar ck_{w}}.
\]

The electron distribution is given by the sum of the initial electron
distribution $f_{0}$ and a perturbation due to the undulator:

\[
f=f_{0}+\frac{N_{phot}}{N_{e}}S
\]

where $S$ is fitted by:

\[
S=\frac{3}{\gamma_{max}}(\frac{\zeta^{2}}{\gamma_{max}^{2}}-\frac{\zeta}{\gamma_{max}}+\frac{1}{2}).
\]

with $\gamma_{max}=2\gamma_{e}^{2}\hbar ck_{w}/(mc^{2})$

Since we suppose that $N_{phot}\ll N_{e}$, we can assume that $N_{e}=\int fd\gamma\approx\int f_{0}d\gamma.$

The energy spread induced by the spontaneous emission is therefore:

\[
\Delta\gamma^{2}=<\gamma^{2}>\approx\frac{2}{3}e^{2}\frac{a_{w}^{2}L_{w}k_{w}^{2}}{\hbar ck_{w}}\int d\gamma S(\gamma)\gamma^{2}
\]

\[
=\frac{14}{15}e^{2}\frac{a_{w}^{2}L_{w}k_{w}^{3}}{(m^{2}c^{3})}\gamma_{e}^{4}\hbar.
\]

Introducing the Compton wavelength $\lambda_{c}=h/(mc)=2.4$$10^{-12}$m,
we obtain:

\[
\Delta\gamma^{2}=\frac{14}{15}r_{e}\frac{\lambda_{c}}{2\pi}a_{w}^{2}L_{w}k_{w}^{3}\gamma_{e}^{4}
\]
 in agreement with the result of Ref {[}LCLS,{]}.

If this relative bandwidth enlargement within the first gain length
is less than the natural FEL bandwidth, the coherent emission can
occur. This can be translated into a condition for the current density:
\[
\frac{\Delta\gamma^{2}}{\gamma^{2}}=\frac{14}{15}r_{e}\frac{\lambda_{c}}{2\pi}a_{w}^{2}L_{w}k_{w}^{3}\gamma_{e}^{2}<\rho^{2}
\]

or:

\[
J>3.5e-20\gamma_{e}^{5}/\lambda_{w}^{4}
\]

\end{document}